%% file: IEEE-conference-template-062824.tex
\documentclass[conference]{IEEEtran}
\IEEEoverridecommandlockouts

\usepackage{cite}
\usepackage{amsmath,amssymb,amsfonts}
\usepackage{algorithmic}
\usepackage{graphicx}
\usepackage{textcomp}
\usepackage{xcolor}
\usepackage{url}

\input{macros}

\begin{document}

\title{SSCBench: Evaluating the Evidential Validity of Fault-Injection Tests for Tool-Using LLM Agents}

\author{\IEEEauthorblockN{Xincheng He\IEEEauthorrefmark{1}\IEEEauthorrefmark{2}, Wanli Dong\IEEEauthorrefmark{1}, Zhaoqiang Guo\IEEEauthorrefmark{3},Yan Liu\IEEEauthorrefmark{4}\IEEEauthorrefmark{1}, Lei Xu\IEEEauthorrefmark{2}
}

\IEEEauthorblockA{\IEEEauthorrefmark{1} School of Artificial Intelligence and Computer Science, Jiangnan University, China}
\IEEEauthorblockA{\IEEEauthorrefmark{2} State Key Laboratory for Novel Software Technology, Nanjing University, China} 
\IEEEauthorblockA{\IEEEauthorrefmark{3} College of Computer Science and Technology, Zhejiang University, China}
\IEEEauthorblockA{\IEEEauthorrefmark{4} School of Internet of Things Engineering, Wuxi University of Technology, China}
}

\maketitle

\input{sections/00_abstract}


\input{sections/01_introduction}
\input{sections/02_background}
\input{sections/03_stf_model}

\input{sections/04_sscbench}

\input{sections/05_study_design}

\input{sections/06_results}

\input{sections/07_discussion}
\input{sections/08_threats}
\input{sections/09_related}
\input{sections/10_conclusion}

\bibliographystyle{IEEEtran}
\bibliography{refs}

\end{document}

%% file: macros.tex
\usepackage{booktabs}
\usepackage{array}
\newcolumntype{L}[1]{>{\raggedright\arraybackslash}p{#1}}
\usepackage{multirow}
\usepackage{makecell}
\usepackage{xspace}
\usepackage{pifont}
\usepackage{enumitem}
\usepackage{subcaption}
\usepackage[most]{tcolorbox}
\usepackage{tikz}
\usetikzlibrary{arrows.meta,positioning,shapes.geometric,fit,calc,decorations.pathreplacing}

\newcommand{\bench}{\textsc{SSCBench}\xspace}          
\newcommand{\taubench}{$\tau$-bench\xspace}

\newcommand{\op}[1]{\textsf{#1}\xspace}                

\newcommand{\vorig}{\ensuremath{v}}
\newcommand{\vmut}{\ensuremath{v'}}

\newcounter{finding}
\newtcolorbox{findingbox}{colback=gray!6,colframe=black!55,boxrule=0.5pt,
  arc=1pt,left=4pt,right=4pt,top=2pt,bottom=2pt,fontupper=\small}
\newcommand{\finding}[1]{%
  \refstepcounter{finding}%
  \begin{findingbox}\textbf{Finding \thefinding.} #1\end{findingbox}}

\newcounter{guideline}
\newtcolorbox{guidelinebox}{colback=white,colframe=black!55,boxrule=0.5pt,
  arc=1pt,left=4pt,right=4pt,top=2pt,bottom=2pt,fontupper=\small}
\newcommand{\guideline}[1]{%
  \refstepcounter{guideline}%
  \begin{guidelinebox}\textbf{Recommend \theguideline.} #1\end{guidelinebox}}

\newtcolorbox{answerbox}{colback=white,colframe=black,boxrule=0.8pt,
  arc=1pt,left=4pt,right=4pt,top=2pt,bottom=2pt,fontupper=\small}

\newtcolorbox{insightbox}{colback=white,colframe=black!40,boxrule=0.4pt,
  leftrule=2.5pt,arc=0pt,left=4pt,right=4pt,top=2pt,bottom=2pt,fontupper=\small}

\newcommand{\rqone}{\textbf{RQ1}\xspace}
\newcommand{\rqtwo}{\textbf{RQ2}\xspace}
\newcommand{\rqthree}{\textbf{RQ3}\xspace}
\newcommand{\rqfour}{\textbf{RQ4}\xspace}

%% file: sections/00_abstract.tex
\begin{abstract}
Fault injection is increasingly used to evaluate the reliability of tool-using LLM agents. However, there has been limited study of how fault-adoption results should be interpreted when the agent itself determines which authoritative observations become visible during execution. In this paper, we present a systematic study of this evidential validity problem in agent fault-injection evaluation. We develop a measurement protocol that specifies what observations can refute an injected assertion, determines whether they can become visible before the affected fact is first used, and records whether the evaluated execution actually realizes this condition. We construct SSCBench as an instantiation of the protocol and evaluate four fault operators and five agent configurations over 1,191 faulted executions in two \taubench environments. Our experiments show that an admitted fault case and agent configuration can realize substantially different evidential conditions across executions, and that aggregate adoption can remain well defined even when the population supporting a timely-counterevidence claim is sparse or absent. For example, among 44 adopted runs in which counterevidence eventually became visible, only 17 received it before first use, while 27 received it afterward. We also find that first-error timing and later stance revision need not coincide, and that automated trajectory analysis can recover adoption without reliably recovering the first faulty-reliance event needed for temporal diagnosis. We argue that the evidential condition realized by an execution and the population supporting a claim-specific interpretation are part of fault-injection evaluation itself and should be reported before adoption is interpreted as failure under pre-use counterevidence.

\end{abstract}

%% file: sections/01_introduction.tex
\section{Introduction}
\label{sec:intro}

Fault injection~\cite{hsueh1997fault} is increasingly used to evaluate the reliability of
tool-using LLM agents. These agents interact with software systems through
APIs that read and modify persistent state~\cite{yao2023react,schick2023toolformer,li2023apibank,qin2024toolllm}, making controlled faults at tool
interactions a natural way to study how erroneous information propagates
through an execution. Existing evaluations report outcomes such as task
degradation, recovery, and the propagation of faulty information into later
actions or responses
\cite{agentchaos,toolbenchx,toolmaze,agentnoisebench,agentstrusttools,agentprop}.

While substantial effort has gone into constructing faults and studying their
downstream effects, a basic question remains when interpreting the resulting
adoption outcomes. A claim that an agent relied on an injected fault despite
contradicting evidence requires more than fault propagation: valid
counterevidence must have become visible before the agent first used the
affected fact. Evaluating this condition poses two particular challenges for
tool-using agents. First, the evaluator must determine what observation
actually refutes the designated injected assertion. Second, because the agent
selects subsequent tool interactions, the evaluator must distinguish whether
such refutation is available through the interface from whether it becomes
visible before use in the realized execution.

The first issue concerns the specification of counterevidence. Retrieving
information related to an affected fact does not necessarily establish a
refutation of the injected assertion. For example, in our \op{ItemOption}
operator, an authoritative product record lists valid product variants, but it
refutes the corrupted option of a particular order item only after the visible
execution establishes which product variant corresponds to that item.
Treating every relevant read as counterevidence can therefore assign an
evidential condition that the execution has not yet established.

A second issue arises from the observation path taken during execution. The
evaluation interface may provide a normal interaction through which an
injected assertion can be checked, but the agent determines whether and when
that interaction occurs. Consider a reservation update that is suppressed
while the tool still returns \texttt{status: success}. An authoritative
read-back of the unchanged reservation can refute this acknowledgement. One
execution may perform that read before relying on the reported success,
another may perform it only afterward, and another may never perform it.
Thus, the opportunity to refute a fault and the evidential condition realized
by an execution are different properties.

Another issue concerns how an adoption outcome is interpreted~\cite{klees2018evaluating,boehme2022reliability,kim2024directedfuzzers}. Aggregate
adoption establishes that an injected assertion propagated into faulty
reliance, but it does not by itself establish that this happened after timely
counterevidence. Likewise, eventual counterevidence combines executions in
which the evidence arrived before use with executions in which it arrived only
afterward. A claim about reliance despite timely counterevidence therefore
requires a different observed population from a claim about fault propagation
alone. We refer to this requirement as the \emph{evidential validity} of a
fault-injection claim.

Finally, these execution-level conditions must be recovered reliably from
agent trajectories. Generic failure localization or anomaly detection may
identify that something went wrong without determining whether the agent
relied on the designated injected assertion or when that reliance first
occurred. An automated analysis may therefore recover a downstream fault
signal while disagreeing with the semantic and temporal events required to
interpret the fault-injection result.

In this paper, we systematically study these issues in fault-injection
evaluation of tool-using agents. We develop a measurement protocol that
specifies the counterevidence for each injected assertion, establishes whether
a legal continuation from the fixed faulted context can expose that
counterevidence before the affected fact is first used, and reconstructs
whether the evaluated execution actually realizes this condition. We construct
\bench to instantiate the protocol on the retail and airline environments of
\taubench~\cite{yao2024taubench}. \bench contains four fault operators spanning
three counterevidence-acquisition structures. Our experiments cover five agent
configurations and 1,191 faulted executions. We study how the realized
evidential condition varies across executions, how it changes the empirical
support behind adoption outcomes, how first faulty reliance relates to the
primary use anchor, and how accurately automated trajectory analyses recover
the required measurement events.

Our contributions are summarized as follows:

\begin{itemize}

\item We identify and formalize an evidential-validity condition in
fault-injection evaluation of tool-using agents. Interpreting adoption as
faulty reliance under timely counterevidence requires both a valid refuting
observation for the injected assertion and evidence that this observation
became visible before the affected fact was first used.

\item We develop a measurement protocol and construct \bench to operationalize
this condition. The protocol combines fault-specific counterevidence,
case-level pre-use revealability, and execution-level reconstruction of
counterevidence arrival and dependent use across four fault operators and
three acquisition structures.

\item We empirically demonstrate that these measurement distinctions
materially affect fault-injection evaluation across five agent
configurations and 1,191 faulted executions. Admitted cases do not fix the
evidential condition realized by an execution, aggregate outcomes can differ
from the population supporting a claim-specific interpretation, and generic
trajectory diagnostics can mismatch the semantic and temporal events needed
for measurement. We derive practical guidelines for reporting and
interpreting fault-injection results.

\end{itemize}

%% file: sections/02_background.tex
\section{Fault-Injection Evaluation of Tool-Using Agents}
\label{sec:background}

\subsection{Evaluation Setting}
\label{sec:bg-process}

Fault injection evaluates how an agent behaves when a controlled error is
introduced during its interaction with the environment
\cite{agentchaos,toolbenchx,toolmaze,ruan2024toolemu,debenedetti2024agentdojo}. A typical evaluation selects a task and
agent configuration~\cite{just2014defects4j,metzman2021fuzzbench,liu2024agentbench,zhou2024webarena,xie2024osworld}, applies an intervention at an eligible tool interaction~\cite{zhang2022fixreverter},
records the resulting trajectory~\cite{ma2024agentboard}, and derives a downstream outcome such as
task success, recovery, or propagation of faulty information into a later
action or response~\cite{agentstrusttools,agentprop}.

For tool-using agents, the trajectory also determines what information becomes
available after the intervention. Subsequent tool calls may expose
authoritative state that contradicts the injected error, expose it only after
the error has already influenced the execution, or never expose it at all.
Thus, two executions can contain the same injected fault and the same
downstream outcome while differing in the information available when that
outcome is produced.

\begin{figure*}[t]
\centering
\includegraphics[width=0.8\linewidth]{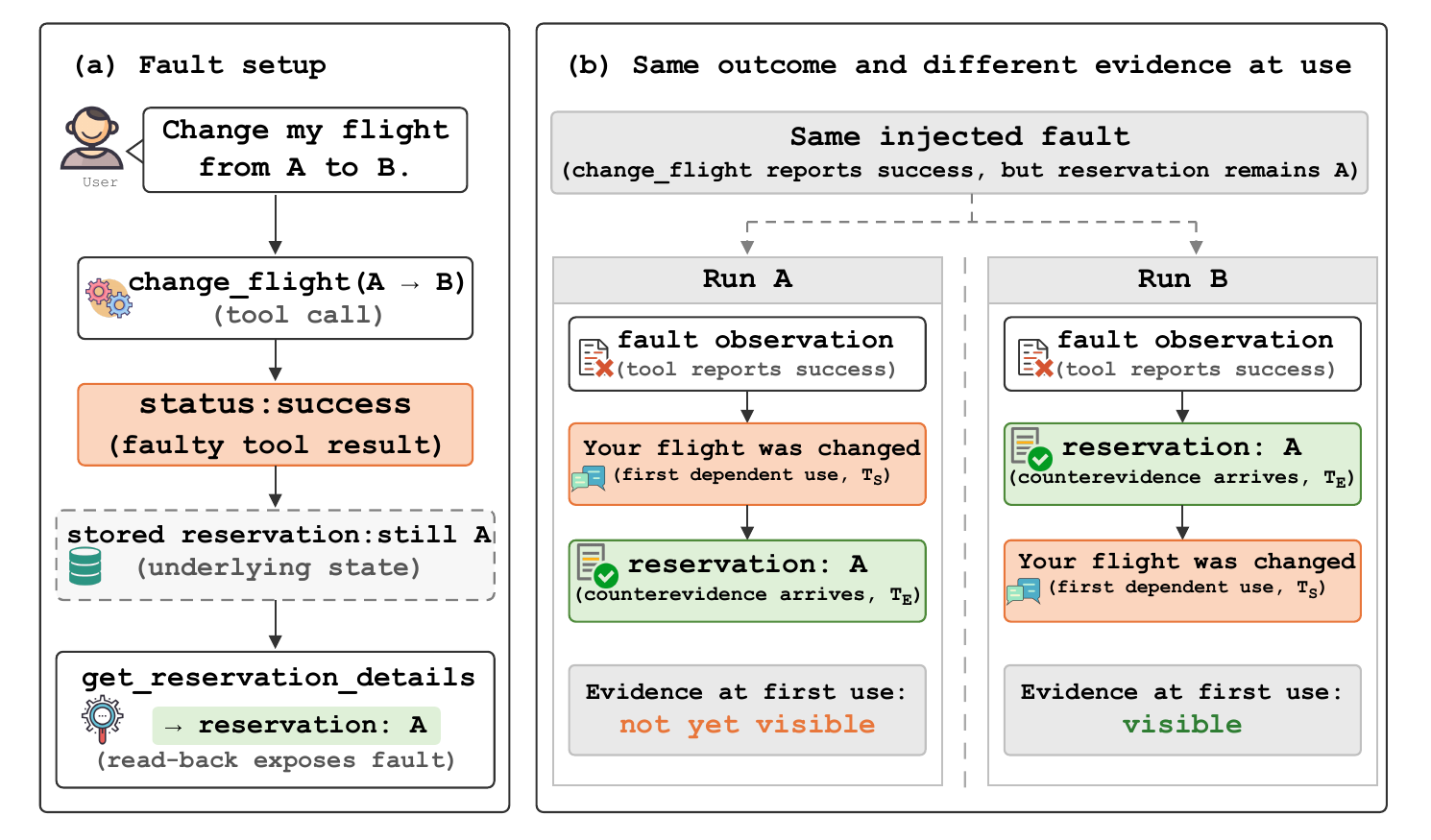}
\caption{A motivating example with same outcome and different evidence.}
\label{fig:motivating}
\vspace{-5pt}
\end{figure*}

\subsection{Motivating Example}
\label{sec:bg-example}

Consider the airline task in Figure~\ref{fig:motivating}, adapted from
\taubench~\cite{yao2024taubench}. The user asks the agent to change an existing
reservation. We inject a \op{WriteEffect} fault that suppresses the requested
update while preserving the tool's successful acknowledgement. The tool
returns \texttt{status: success}, although the stored reservation remains
unchanged. A later \texttt{get\_reservation\_details} call can expose the
unchanged reservation and contradict the reported success.
Figure~\ref{fig:motivating}b shows two executions of this same fault. In
Run~A, the agent tells the user that the flight was changed and reads the
unchanged reservation only afterward. In Run~B, the unchanged reservation
becomes visible before the same claim is made. Both executions
produce the same faulty downstream claim, but they are under different
evidential conditions. In the figure, $T_E$ marks the arrival of the
contradicting observation and $T_S$ marks the first use of the affected
information.

The example exposes two distinctions that are important for fault-injection
evaluation. First, information related to a fault is not necessarily sufficient
to refute it. For example, in an item-option error, a product record becomes
refuting evidence only after the execution establishes that the record
corresponds to the affected order item. Second, the availability of a
refuting observation through the interface does not determine whether a
particular execution obtains it before the faulty information is used. The
agent's subsequent interactions determine that realization.
These distinctions affect how a downstream adoption outcome can be
interpreted. Observing that an injected error propagated into an agent action
or response establishes fault propagation. Interpreting the same outcome as
reliance despite timely counterevidence additionally requires evidence that
the relevant refutation had become visible before use. 

%% file: sections/03_stf_model.tex
\section{Evidential Conditions for Fault-Injection Evaluation}
\label{sec:estimand}

\subsection{Measurement Objects}
\label{sec:defs}
\label{sec:arrival}

\textit{Definition 3.1 (Injected fault).}
An \emph{injected fault} is an intervention at a tool interaction that causes
the interaction to convey a faulty assertion $\vmut$ about an affected fact
$f$, while the authoritative environment state supports a different assertion
$\vorig$. The observation carrying $\vmut$ is the
\emph{fault-bearing observation}. The affected fact is scoped to a target entity, attribute, and relevant
state. For a state-changing operation, it may instead concern the claimed
effect of a designated logical request. A run is \emph{triggered} when the
intervention takes effect and the fault-bearing observation is delivered to
the agent.

\textit{Definition 3.2 (Dependent use).}
A \emph{dependent use} of $f$ is an agent message, tool call, or handoff whose
observable content or arguments depend on a value or proposition about $f$,
including a derived value. Correct, qualified, and faulty uses all count.
A call that only retrieves information about $f$ is not a dependent use
unless its arguments themselves depend on $f$.

\textit{Definition 3.3 (Faulty reliance and adoption.)}
A \emph{faulty reliance} is a dependent use that treats $\vmut$, or a value
derived from it, as the operative task fact. A run \emph{adopts} the fault,
written $A=1$, if it contains at least one faulty reliance. Using a disputed value only to verify that value does not constitute faulty
reliance.

\textit{Definition 3.4 (Valid counterevidence.)}
An observation, or set of observations, is \emph{valid counterevidence} to
$\vmut$ if all of the following conditions hold.

\begin{itemize}[leftmargin=*,topsep=2pt,itemsep=1pt]
\item \emph{Authority}: its source is at least as authoritative for $f$ as the
fault-bearing observation.
\item \emph{Entity}: it concerns the same target entity.
\item \emph{Scope}: it concerns the state or logical request named by the
faulty assertion.
\item \emph{Independence}: it is grounded in information that the intervention
did not alter to agree with $\vmut$.
\item \emph{Sufficiency}: together with task and tool semantics already
available to the agent, it is sufficient to contradict $\vmut$.
\end{itemize}

Evaluator-only knowledge does not contribute to sufficiency. Independence
concerns the information source rather than the message boundary; valid
counterevidence may therefore be co-delivered with the fault-bearing
observation. Agent-generated assertions do not create counterevidence;
counterevidence must be grounded in task-authoritative information delivered
through the evaluation interface.

The \emph{model-visible history} contains the user and agent messages, tool
calls, and tool observations available in the agent context. Evaluator-only
environment state and intervention logs are excluded.

Let $T_F$ denote delivery of the fault-bearing observation and $T_E$ the
earliest step by which valid counterevidence has fully entered the
model-visible history. If several observations are jointly required,
$T_E$ is the step at which the last required observation arrives.
Evidence already visible before fault delivery may yield $T_E<T_F$.
If no valid counterevidence appears in the observed execution,
$T_E=\infty$; if arrival cannot be determined, $T_E$ is undecidable.

The first post-fault dependent use and first faulty reliance are
\begin{equation}
\label{eq:anchors}
\begin{aligned}
T_S &= \min \{t>T_F: \operatorname{Use}_f(e_t) \}, \\
T_U &= \min \{t>T_F: \operatorname{FaultyReliance}_f(e_t) \}.
\end{aligned}
\end{equation}

$T_S$ is defined independently of adoption and serves as the primary temporal
anchor. $T_U$ is a secondary diagnostic anchor for locating the first faulty
reliance. Whenever both are decidable,
$ T_F<T_S\leq T_U.$
Adoption may be decidable even when $T_U$ cannot be localized. A run with no
post-fault dependent use is distinct from a run whose $T_S$ is undecidable.

\subsection{Pre-Use Revealability}
\label{sec:revealability}

A valid fault conveys a task-relevant faulty assertion while preserving the
non-target semantics required by the intervention.

\textit{Definition 3.5 (Pre-use revealability.)}
For a valid candidate fault $c$, let $h_c$ denote its fixed injection context
and let $\rho$ denote a legal continuation under the same faulted environment,
interface, and permissions. The candidate is \emph{pre-use revealable} if
\begin{equation}
\label{eq:revealability}
\exists \rho
\quad\text{such that}\quad
T_E(h_c\Vert\rho)<T_S(h_c\Vert\rho)<\infty.
\end{equation}

The injection context fixes the target entity, injection point, model-visible
history, fault parameters, intervention persistence, interface, and
permissions. Every interaction in $\rho$ must use arguments obtainable from
the model-visible history and the task interface, without evaluator-only
information.

A fault is \emph{admitted} when it is valid and pre-use revealable.
Revealability is a property of the fault context and evaluation interface:
it establishes that valid counterevidence can become model-visible before
first use through normal interaction. Whether that opportunity is realized
is a property of an individual execution.

\subsection{Realized Evidential Condition}
\label{sec:strata}
\label{sec:outcomes}
\label{sec:identity}
\label{sec:diagnostic}

For an admitted and triggered run with a decidable first dependent use, the
\emph{evidential condition} is defined by the ordering of $T_E$ and $T_S$:
\begin{equation}
\label{eq:strata}
Z_S=
\begin{cases}
\mathrm{pre},    & T_E<T_S,\\
\mathrm{late},   & T_S<T_E<\infty,\\
\mathrm{absent}, & T_E=\infty.
\end{cases}
\end{equation}

A pre-arrival run has valid counterevidence visible before first use.
A late-arrival run obtains it only after first use. An absent-arrival run
contains no valid counterevidence in the observed model-visible history.
An undecidable arrival is not classified as absent, and a run with no
dependent use receives no arrival-relative-to-use classification.

Revealability and arrival describe different properties. Revealability
establishes a pre-use refutation opportunity for the case; $Z_S$ records the
condition realized by a particular execution. The same admitted fault can
therefore produce different values of $Z_S$.

The downstream outcome and the timing of counterevidence support distinct
claims:
\begin{align}
A=1
&\qquad \text{fault adoption},\\
A=1 \land T_E<\infty
&\qquad \text{adoption with eventual counterevidence},\\
A=1 \land T_E<T_S
&\qquad \text{adoption with pre-use counterevidence}.
\end{align}

For adopted runs, $T_E<T_U$ supports a complementary reliance-specific
claim: counterevidence was visible before the first faulty reliance.
We use $T_S$ as the common stratification anchor because it is defined
independently of adoption and therefore applies uniformly to adopted and
non-adopted runs. $T_U$ instead localizes the first error within adopted
runs.
Arrival records when valid counterevidence enters the model-visible history.
Accordingly, $T_E<T_S$ and $T_E<T_U$ characterize different temporal
relations. A run with $T_S<T_E<T_U$ is late relative to first dependent use
but pre-reliance relative to the first faulty reliance.

%% file: sections/04_sscbench.tex
\section{\bench: Operationalizing Arrival-Aware Fault Injection}
\label{sec:sscbench}

\begin{figure*}[h]
\centering
{\includegraphics[width=\linewidth]{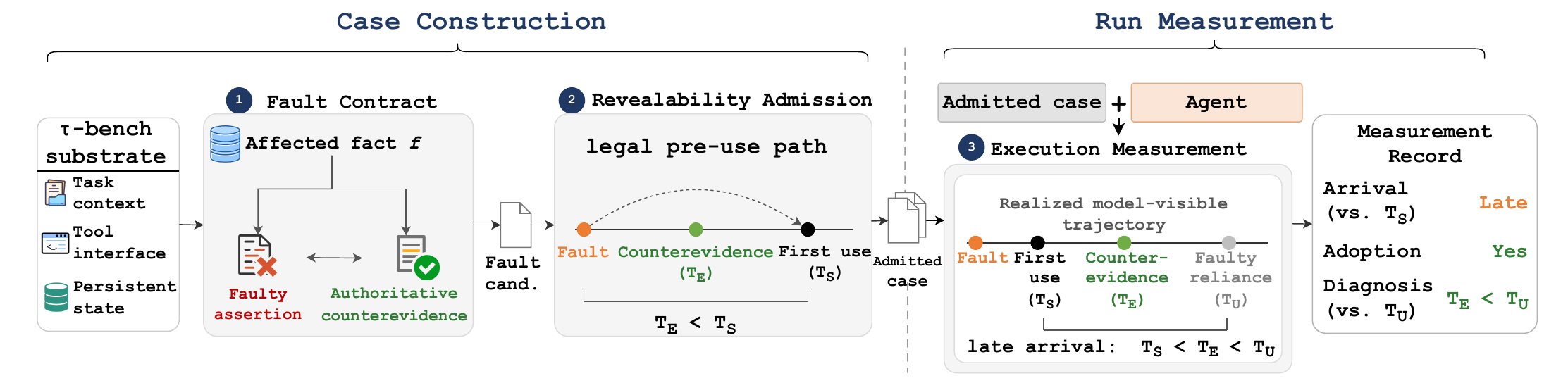}}
\caption{Overview of \bench.}
\label{fig:sscbench-pipeline}
\end{figure*}

\bench is a fault-injection benchmark in which every admitted case carries
an explicit relation among a faulty assertion, the observations that can
refute it, and the execution events that determine when those observations
become model-visible. Figure~\ref{fig:sscbench-pipeline} summarizes its three
mechanisms. A \emph{fault contract} binds an intervention to the affected fact
and its valid counterevidence. \emph{Revealability admission} establishes a
legal continuation through which that counterevidence can arrive before the
first dependent use. During evaluation, \emph{execution reconstruction}
recovers the realized fault, evidence, and use events from the trajectory.
These mechanisms make counterevidence arrival both a controlled
property of benchmark construction and an observable property of execution.

We instantiate \bench in the retail and airline environments of \taubench.
The source environments provide task obligations that scope the affected
fact, tool interfaces that delimit legal interventions and observations, and
persistent records that can serve as authoritative evidence. \bench augments
this substrate with a scoped fault relation, an explicit counterevidence
relation, and trace records that preserve the temporal events required for
execution-level measurement.

\subsection{Fault-Contract Instantiation}
\label{sec:operators}

The unit of case construction in \bench is a \emph{fault contract}.
A fault contract gives an injected mutation its measurement semantics by
relating a faulty assertion about an affected fact to the independent
observations that can refute it. Specifically, the contract fixes the
target entity and state scope, the injected assertion, the authoritative
basis for counterevidence, and the semantic relation that makes the
contradiction sufficient. It also determines how such counterevidence can
be acquired through the task interface.

We derive operator templates from task obligations and tool contracts and
instantiate them over concrete environment records. The three read-value
operators modify a designated field in a returned observation while leaving
the corresponding authoritative record unchanged. \op{WriteEffect} instead
intervenes on a state transition and suppresses a matched update while
preserving the successful acknowledgement returned by the tool. Each
instantiation binds the intervention to a target entity, a concrete affected
fact, and the authoritative observations that can contradict the resulting
faulty assertion.

The four operators differ in their injected faults, the relations that make
counterevidence sufficient, and the routes by which it can become
model-visible. \op{ReturnDate} moves a reservation return date earlier,
\op{PaySource} flips a payment category, and \op{ItemOption} substitutes
an alternative option for an ordered item.

The operators instantiate three counterevidence-acquisition structures.
For \op{ReturnDate} and \op{PaySource}, the observations required for
refutation are co-delivered with the fault-bearing observation. Their
sufficiency nevertheless depends on agent-visible semantics that connect
the retained field to the affected fact. For \op{ReturnDate}, the retained
outbound leg becomes counterevidence through the ordering constraint of a
round trip. For \op{PaySource}, the retained payment identifier becomes
sufficient only when its relation to the payment category is established
to the agent; an empirical naming pattern in stored data alone does not
supply this semantics.

\op{ItemOption} instead uses a cross-record refutation relation. The
authoritative product record identifies the valid variants, but it becomes
counterevidence for a particular corrupted order item only after the product
variant is bound to that item. The required product record may already be
present in the visible prefix or may enter the execution through a later
lookup. Counterevidence availability and counterevidence arrival can
therefore differ across executions of the same instantiated fault.

\op{WriteEffect} moves the refutation relation across a state transition.
Its faulty assertion is carried by a successful acknowledgement whose tool
contract states that the requested update has been committed. The mutation
suppresses the designated state change while preserving that acknowledgement
and the interface's normal validation. Counterevidence is then provided by
an authoritative read of the persistent reservation state after the
attempt. The read refutes the acknowledgement only while it remains
attributable to the same logical request.

To preserve this attribution, \op{WriteEffect} matches the target request by
reservation, flight changes, cabin, and payment method. Repeated attempts
belong to the same logical request only when no intervening write changes
the affected field. A later reservation read therefore supplies
counterevidence only when its state scope still refers to the suppressed
attempt.

These acquisition structures constrain how valid counterevidence can enter
an execution without fixing whether or when it actually does so. The
resulting fault contract defines the semantic relation required for
measurement.

\subsection{Pre-Use Revealability Admission}
\label{sec:admission}
\label{sec:funnel}

A fault contract determines what constitutes valid counterevidence.
Admission additionally requires the normal interaction surface to permit
that evidence to become model-visible before dependent use. \bench therefore
admits a concrete injection context only when it has a pre-use revealability
witness.

For a candidate $c$, the injection context $h_c$ fixes the model-visible
prefix through the fault-bearing observation together with the target, fault
parameters, intervention persistence, interface, and permissions. A witness
is a legal continuation $\rho$ under the same faulted environment whose tool
arguments are obtainable from that execution and whose delivered
observations satisfy the contract's counterevidence relation before the
first dependent use. The admission condition is$ T_E(h_c\Vert\rho) < T_S(h_c\Vert\rho) < \infty$.
The witness need not complete the task, but it must arise through the normal
task interface under the original permissions. A clean execution cannot
serve as the witness because it does not establish the counterevidence
relation for the injected fault.

An admissible witness presupposes a valid fault contract. The designated
intervention must preserve the operator's non-target semantics, including
schema and identifier constraints, while the affected fact remains relevant
to the task. Evidence authority and sufficiency are grounded in domain and
interface semantics available to the agent rather than in empirical
correlations observed by the evaluator. For example, equality between an
order price and a current catalog price does not establish a semantic
requirement that the two values agree.

The witness trace must establish the complete relation represented by the
admission condition. The trace records the intervention and fault delivery,
the provenance of lookup arguments, the observations that form
counterevidence, and the proposed first dependent use. The evidence must
satisfy authority, entity, scope, independence, and agent-visible
sufficiency, with no earlier dependent use in the visible prefix. We retain
the trace together with the prefix identifier and hash and the witness
continuation, so the admission relation remains tied to the concrete
injection context rather than inferred from the operator alone.

Revealability is context-specific. A witness can transfer to another visible
prefix only when the identifiers, permissions, and refutation relation on
which it depends remain valid; otherwise the new context requires its own
witness. Conversely, absence of a qualifying witness leaves a candidate
unresolved rather than establishing unrevealability. A candidate is admitted
only when both its fault contract and a witness for the applicable injection
context satisfy these conditions. Study inclusion thus operates at the case
level: only candidates with a valid fault contract and applicable pre-use
witness contribute executions to the primary analysis.

\subsection{Execution-Level Event Reconstruction}
\label{sec:ledger}
\label{sec:oracle}

Admission establishes that timely counterevidence is possible for a case.
The evaluated trajectory determines whether that evidence actually arrives
and where its arrival falls relative to downstream use. \bench therefore
reconstructs the measurement events from the model-visible execution rather
than assigning an arrival condition from the operator or its admission
witness.

The execution record is an ordered ledger over the complete model-visible
message sequence. Each tool observation is paired with the call that
produced it, while call arguments, returned content, and agent messages
retain their original delivery order. $T_F$ is located only when both the
designated intervention and delivery of its fault-bearing observation are
established. Environment state used to verify the intervention remains
separate from the model-visible evidence stream: hidden state can establish
that the mutation occurred, but only delivered observations can establish
counterevidence arrival.

\bench recovers $T_E$ by applying the case contract to the delivered
history. A tool name, database value, or evaluator-visible discrepancy alone
is insufficient. For \op{ItemOption}, the affected order item must be bound
to the corresponding product variant before the product record can refute
its option. When the product record appears before that binding, the later
binding completes the counterevidence relation. For \op{WriteEffect}, the
authoritative reservation read must follow the suppressed attempt and remain
scoped to that attempt despite any intervening write. In general, $T_E$ is
the earliest delivery step at which the required observations and the
agent-visible semantic basis jointly suffice to contradict the faulty
assertion. Evidence already present in the visible prefix may therefore
yield $T_E<T_F$. Agent-generated statements do not constitute
counterevidence.

To recover agent-side use events, \bench separates candidate extraction from
semantic classification. An extraction pass identifies post-fault references
to the affected entity and value, including derived values, without
providing the mutation's correctness label. Each candidate event is then
interpreted under the fault contract to determine semantic dependence and
faulty reliance. A search argument derived from the faulty value, for
example, is a dependent use, but it becomes a faulty reliance only when the
surrounding visible context treats that value as an operative task fact
rather than as a value being checked or enumerated. The first determinate
dependent use defines $T_S$; in an adopting run, the first determinate faulty
reliance defines $T_U$. A textual qualification does not cancel an action
that operationally commits to the faulty assertion.

\bench treats decidability as part of the measurement record. A complete
trace with no dependent use is labeled \emph{no-use}; a potentially earlier
but unresolved use instead leaves $T_S$ undecidable. Likewise, one certain
faulty reliance establishes adoption even when an earlier ambiguous event
leaves $T_U$ unknown, whereas the absence of a certain faulty reliance does
not establish non-adoption while a potential use remains unresolved. For
adopted runs, \bench records the last determinate, unreplaced
explicit stance toward the faulty assertion; an erroneous action without a
communicated stance leaves it undecidable. The resulting record stores both
the value and the decidability of $T_E$, $T_S$, $T_U$, adoption, arrival
stratum, and terminal stance.

%% file: sections/05_study_design.tex
\section{Empirical Study Design}
\label{sec:design}

\subsection{Research Questions}
\label{sec:design-rqs}

\begin{itemize}[leftmargin=*,topsep=2pt,itemsep=2pt]

\item \rqone (\emph{Arrival}).
\emph{How often does counterevidence arrive before the first dependent use
of the affected fact?}

\item \rqtwo (\emph{Outcome}).
\emph{How does fault adoption vary across counterevidence arrival
conditions?}

\item \rqthree (\emph{Diagnosis}).
\emph{Among adopted runs, when does counterevidence arrive relative to the
first faulty reliance?}

\item \rqfour (\emph{Measurement Recoverability}).
\emph{How accurately can automated trajectory oracles recover fault
adoption and localize the first faulty reliance?}

\end{itemize}

\subsection{Experimental Setup}
\label{sec:design-subjects}
\label{sec:design-sampling}
\label{sec:design-data}

\textbf{Study Subjects and Sampling.}
We use the retail and airline environments of \taubench.
\op{ItemOption} targets retail order items, while \op{ReturnDate},
\op{PaySource}, and \op{WriteEffect} target airline reservations.
Candidate frames are constructed from task specifications and environment
records before evaluated-agent runs. They contain 105 \op{ReturnDate},
203 \op{ItemOption}, 120 \op{PaySource}, and 174 \op{WriteEffect}
candidates. For each operator, we fix a candidate order using stable identifiers and
screen candidates until 30 cases satisfy fault validity and pre-use
revealability. Sampling cycles through target-entity clusters and does not
use outcomes from the evaluated agents. Entity clusters are reservations for
\op{ReturnDate} and \op{WriteEffect}, order items for \op{ItemOption}, and
users for \op{PaySource}. The final study contains 120 admitted seeds,
30 per operator. Specifically, the \op{ItemOption} sample contains 30 items from 15 orders and
14 tasks. 16 initially selected candidates were replaced after the
final semantic review using the same frozen candidate order, completing
the 30-seed sample by rank 84. Ten admitted cases are target-value controls.
They are retained in the arrival analysis and excluded from the primary
adoption and diagnosis analyses.

\textbf{Agent Configurations and Execution.}
We evaluate five model configurations from three providers.
Table~\ref{tab:configs} reports the model, decoding setting, planned
repetitions, and realized execution scale. F, V, and L use three
pre-numbered repetitions (r0--r2), while T and S use one (r0).
All configurations share the same task scaffold, system prompt, tool schema,
environment, and native \taubench\ user simulator. The simulator uses
\texttt{deepseek-v4-pro} at $T{=}0$ with thinking disabled.
\bench runs on \taubench\ commit \texttt{59a200c}. Each execution starts
from the same reset state. Step limits are 60 for \op{ReturnDate} and
\op{WriteEffect}, 40 for \op{PaySource}, and 30 for \op{ItemOption}.
A seed is executed under fault injection for a configuration only when that
configuration first solves the corresponding clean task. A scheduled
faulted run that is not observed is treated as missing and is not imputed. The resulting execution matrix contains 1,191 faulted runs.

\begin{table}[t]
\centering
\footnotesize
\setlength{\tabcolsep}{2pt}
\caption{Agent configurations and execution scale.}
\label{tab:configs}
\label{tab:matrix}
\begin{tabular*}{0.95\linewidth}{@{\extracolsep{\fill}}c l l c @{} c c c c r@{}}
\toprule
\multicolumn{4}{@{}l}{\textbf{Configurations}} &
\multicolumn{5}{@{}l}{\textbf{Executions}}\\
\cmidrule(r){1-4}\cmidrule(l){5-9}
ID & Model & Decoding & Reps. & RD & IO & PS & WE & Total\\
\midrule
F & \texttt{deepseek-flash}  & $T{=}0$, off & 3 & 29/88 & 18/54 & 30/90 & 30/90 & 322\\
V & \texttt{deepseek-v4-pro} & $T{=}0$, off & 3 & 30/90 & 19/57 & 30/90 & 30/90 & 327\\
L & \texttt{gpt-5.6-luna}    & default      & 3 & 30/90 & 22/66 & 30/90 & 29/87 & 333\\
T & \texttt{gpt-5.6-terra}   & default      & 1 & 30/30 & 18/18 & 30/30 & 26/25 & 103\\
S & \texttt{claude-sonnet-5} & default      & 1 & 30/30 & 19/19 & 30/30 & 27/27 & 106\\
\bottomrule
\end{tabular*}
\end{table}
\begin{table}[t]
\centering
\footnotesize
\setlength{\tabcolsep}{3pt}
\caption{Trajectory-oracle settings.}
\label{tab:oracles}
\begin{tabular*}{0.95\linewidth}{@{\extracolsep{\fill}}l
  L{0.30\linewidth} L{0.34\linewidth} L{0.18\linewidth}@{}}
\toprule
Setting & Method & Input & Output\\
\midrule
B1 & Schema validator & tool returns, clean schema & anomaly, step\\
\addlinespace[2pt]
B2-L0 & LLM trajectory judge & L0 & verdict, step\\
B2-L1 & LLM trajectory judge & L1 & verdict, step\\
\addlinespace[2pt]
B3-native & AgentRx~\cite{agentrx} & native interface, user instruction & verdict, step\\
B3-SAA & AgentRx & minimal L1 adaptation & verdict, step\\
\addlinespace[2pt]
B4 & Step attribution~\cite{zhang2025whichagent} & L0 & step only\\
\addlinespace[2pt]
B5 & Mined state invariants~\cite{ernst2001daikon} & L-DB & violation, step\\
\bottomrule
\end{tabular*}
\end{table}

\textbf{Analysis.}
The primary analyses use r0 for each clean-qualified
configuration--seed pair. The analysis population $\mathcal{P}$ contains
admitted, triggered runs with a decidable post-fault dependent use; no-use
and undecidable runs are reported separately.
We report the pre-, late-, and absent-arrival shares over runs with decidable
$Z_S$. Adoption is reported among triggered runs ($A_T$), within
$\mathcal{P}$ ($A_{\mathcal P}$), and within each arrival condition
($a_s$). Among adopted runs with decidable $T_E$ and $T_U$, we report
$ q_U=P(T_E<T_U\mid A=1,\ r\in\mathcal{P})$,
together with the number of runs satisfying $T_S<T_E<T_U$.
$Y_{\mathrm{end}}$ records the final decidable stance toward the faulty
assertion for adopted runs.

We report 95\% intervals using Wilson intervals when each contributing
entity cluster provides one evaluable run and cluster bootstrap intervals
otherwise ($B=2{,}000$). Sensitivity analyses use common clean-solved seeds,
all available repetitions, a stricter \op{ItemOption} admission rule, and
the inclusion of target-value controls. For repeated configurations, we also
record whether a seed--configuration pair changes arrival condition across
repetitions.
We additionally re-express the same \op{WriteEffect} r0 runs as adoption,
adoption with eventual counterevidence, and adoption with pre-use
counterevidence. Two paired checks modify either the contradiction relation
(\op{ItemOption} and \op{WriteEffect}, F/V) or the reservation read-back cue
(\op{WriteEffect}, F/V/L/S). Arrival-stratified comparisons are descriptive
rather than causal.

\textbf{Human Measurement Validation.}
Reviewers A, B, and C independently perform the human review, and reviewer D
adjudicates disagreements. At case level, reviewers assess task relevance,
evidence authority, agent-visible sufficiency, and entity or request scope
without access to evaluated-agent outcomes. 
For execution-level validation, A, B, and C first complete five calibration
trajectories and then independently review 120 faulted trajectories sampled
across operators, configurations, and automatic arrival conditions. They
label $T_E$, $T_S$, $T_U$, $A$, and quantity-specific decidability from the
fault contract and numbered model-visible trajectory. Configuration identity,
experimental condition, automatic labels, and automatic event locations are
masked. $Z_S$ is derived from the reviewed $T_E$ and $T_S$.
A label is accepted directly only when A, B, and C agree. Any disagreement
is adjudicated by D; a 2--1 majority is not automatically accepted.
We report pre-adjudication Fleiss' $\kappa$ for categorical quantities and
exact-step agreement when all three reviewers locate the event.
Automatic measurements are compared with the adjudicated reference on event
status, exact event location when both sources locate the event, $Z_S$, and
$A$.

\textbf{Automated Oracle Evaluation.}
The oracle reference set contains the 120 adjudicated faulted trajectories
and 40 clean trajectories, ten per operator, as negative controls.
Adoption evaluation uses faulted trajectories with decidable human $A$;
localization uses adopted faulted trajectories with decidable human $T_U$.
We report adoption precision, recall, and F1, together with false-positive
rates on clean and non-adopting faulted runs. Coverage is reported separately.
Adoption recall uses the full applicable reference set, so abstentions count
as missed positives. Localization is measured by exact and $\pm1$-step
accuracy over the full applicable adopted set, with abstentions counted as
nonmatches. Results are also stratified by operator and human arrival
condition.
Table~\ref{tab:oracles} summarizes the evaluated settings. L0 contains the
numbered trajectory, tool schema, and domain policy; L1 additionally provides
the fault-specific relation contract; L-DB contains invariants mined from
clean database states. LLM-based settings use one call per trajectory and
setting.
B3-native is evaluated on the 40 \op{ItemOption} faulted trajectories for
which the native AgentRx setting was run with one-shot; B3-SAA uses the full faulted
reference set. For B1 and B5, an emitted anomaly or invariant violation is
treated as a positive verdict. B4 produces only a step and is excluded from
adoption metrics.

%% file: sections/06_results.tex
\section{Results}
\label{sec:results}

\subsection{Admission and Measurement Validation}
\label{sec:validity}

Across the
1,191 faulted executions, 1,189 triggered the intervention. All 120 admitted
cases passed the final semantic checks and had a recorded pre-use witness.

Table~\ref{tab:study-funnel} reports the primary-r0 execution funnel and
measurement validation against the 120 adjudicated faulted trajectories.  In panel (a),
$n_Z/u_Z$ gives decidable/undecidable arrival among runs with a first
dependent use. Panel (b) compares frozen automatic measurements with the
adjudicated reference; exact-step agreement applies when both locate the
event.

Before adjudication, A/B/C agreement was $\kappa=.972$ for $A$,
$\kappa=.985$ for $Z_S$, $\kappa=.972$ for $T_U$ status, and
$\kappa=.812$ for $T_S$ status. Exact steps agreed in 84/86 located
$T_U$ events and 114/115 located $T_S$ events. The frozen automatic
measurements matched the adjudicated reference in 66/70 applicable $Z_S$
labels and 111/120 $A$ labels.
We further reviewed all 96 primary-r0 \op{ItemOption} trajectories.
Twenty-one were already in the validation sample; the remaining 75 introduced
no additional disagreement with the frozen automatic measurements. The
primary analyses therefore retain the frozen automatic labels.

\subsection{RQ1: Counterevidence Arrival}
\label{sec:rq1}

The co-delivered operators were uniformly pre-arrival: all 149 evaluable
\op{ReturnDate} runs and all 150 evaluable \op{PaySource} runs received
valid counterevidence before first use. Table~\ref{tab:arrival} reports the
two acquisition-dependent operators, in which $n/k$ gives runs/entities with decidable $Z_S$; the undecidable arrival within $\mathcal P$ is 0.
\op{ItemOption} produced 40 pre-, 49 late-, and two absent-arrival runs
among 91 evaluable executions. \op{WriteEffect} exhibited still wider
variation: L was pre-arrival in 29/29 runs, V absent-arrival in 30/30,
and T late-arrival in 20/25. Common-seed and all-repetition analyses
preserved these differences.
Arrival also varied after fixing both the case and configuration. Among F,
V, and L, 11/89 seed--configuration groups with repeated evaluable executions
crossed arrival conditions. In the matched \op{WriteEffect} read-cue check,
removing the cue reduced read-back from 61/115 to 18/115 and pre-use arrival
from 54/115 to 0/115.

\begin{table*}[t]
\centering
\footnotesize
\setlength{\tabcolsep}{2.7pt}
\caption{Admission and measurement validation.}
\label{tab:study-funnel}
\label{tab:oracle}
\begin{tabular*}{0.95\linewidth}{@{\extracolsep{\fill}}l c c c @{} l c c@{}}
\toprule
\multicolumn{4}{@{}l}{\textbf{(a) Execution funnel}} &
\multicolumn{3}{@{}l}{\textbf{(b) Label validation, 120 runs}}\\
\cmidrule(r){1-4}\cmidrule(l){5-7}
Operator & Triggered & No use/$u_S$ & $n_Z/u_Z$ &
Quantity & Auto/ref. status & Exact step\\
\midrule
\op{ReturnDate} & 149 & 0/0 & 149/0 &
$T_E$ & 70/70 & 57/57\\
\op{ItemOption} & 96 & 5/0 & 91/0 &
$T_S$ & 117/120 & 111/115\\
\op{PaySource} & 150 & 0/0 & 150/0 &
$T_U$ & 111/120 & 75/79\\
\op{WriteEffect} & 141 & 0/0 & 141/0 &
$Z_S$ & 66/70 & --\\
All & 536 & 5/0 & 531/0 &
$A$ & 111/120 & --\\
\bottomrule
\end{tabular*}
\end{table*}

\begin{table*}[t]
\centering
\footnotesize
\setlength{\tabcolsep}{5pt}
\caption{Counterevidence arrival for acquisition-dependent operators.}
\label{tab:arrival}
\newcommand{\arrcell}[2]{#1\,[#2]}
\begin{tabular*}{0.95\textwidth}{@{\extracolsep{\fill}}c r c c c r@{}}
\toprule
ID & $|\mathcal P|$ & $n/k$ &
Pre [95\% CI] & Late [95\% CI] & Absent [95\% CI] \\
\midrule
\multicolumn{6}{@{}l}{\op{ItemOption}}\\
F & 18 & 18/10 & \arrcell{28\%}{0\%,57\%} &
\arrcell{67\%}{38\%,100\%} & \arrcell{6\%}{0\%,21\%}\\
V & 19 & 19/10 & \arrcell{26\%}{0\%,54\%} &
\arrcell{74\%}{46\%,100\%} & \arrcell{0\%}{0\%,0\%} \\
L & 18 & 18/11 & \arrcell{78\%}{57\%,95\%} &
\arrcell{22\%}{5\%,43\%} & \arrcell{0\%}{0\%,0\%} \\
T & 17 & 17/10 & \arrcell{35\%}{13\%,57\%} &
\arrcell{65\%}{43\%,88\%} & \arrcell{0\%}{0\%,0\%} \\
S & 19 & 19/11 & \arrcell{53\%}{20\%,75\%} &
\arrcell{42\%}{20\%,72\%} & \arrcell{5\%}{0\%,19\%}\\
\midrule
\multicolumn{6}{@{}l}{\op{WriteEffect}}\\
F & 30 & 30/30 & \arrcell{83\%}{66\%,93\%} &
\arrcell{3\%}{1\%,17\%} & \arrcell{13\%}{5\%,30\%}\\
V & 30 & 30/30 & \arrcell{0\%}{0\%,11\%} &
\arrcell{0\%}{0\%,11\%} & \arrcell{100\%}{89\%,100\%} \\
L & 29 & 29/29 & \arrcell{100\%}{88\%,100\%} &
\arrcell{0\%}{0\%,12\%} & \arrcell{0\%}{0\%,12\%} \\
T & 25 & 25/25 & \arrcell{8\%}{2\%,25\%} &
\arrcell{80\%}{61\%,91\%} & \arrcell{12\%}{4\%,30\%}\\
S & 27 & 27/27 & \arrcell{4\%}{1\%,18\%} &
\arrcell{22\%}{11\%,41\%} & \arrcell{74\%}{55\%,87\%} \\
\bottomrule
\end{tabular*}
\end{table*}

\finding{\textbf{An admitted fault case did not determine the evidential
condition under which it was exercised.}
Even after fixing the fault contract and agent configuration,
counterevidence exposure varied with the realized observation path and was
sensitive to interface cues that changed whether authoritative state was
revisited before use.}

\subsection{RQ2: Outcomes Across Arrival Conditions}
\label{sec:rq2}

\begin{figure}[t]
\centering
\includegraphics[width=0.95\linewidth]{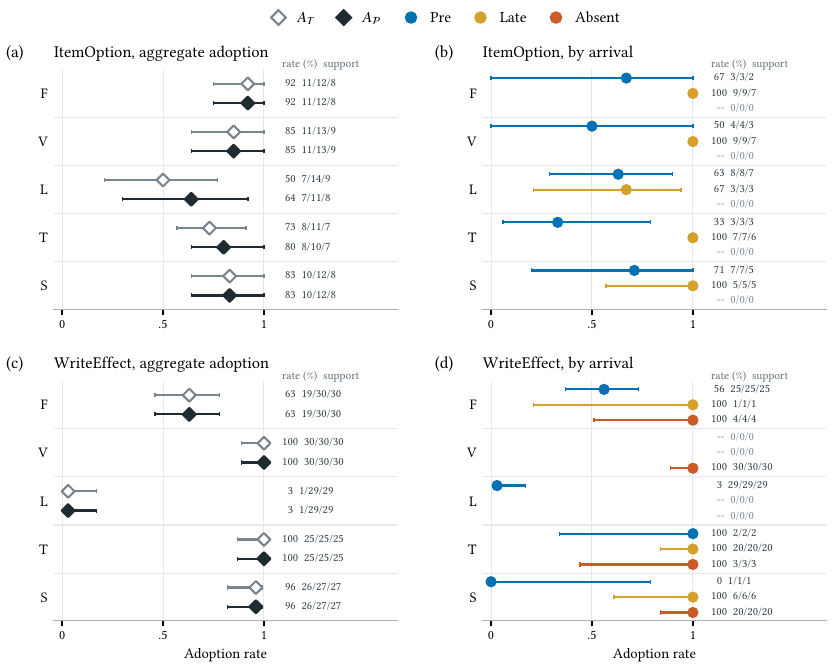}
\caption{Adoption by arrival condition.}
\label{fig:rq2-full-evidence}
\end{figure}

Figure~\ref{fig:rq2-full-evidence} reports aggregate and
arrival-conditioned adoption. The top row shows \op{ItemOption}:
panel (a) compares $A_T$ (open diamonds) and $A_{\mathcal P}$
(filled diamonds), while panel (b) shows adoption in pre, late, and
absent arrival strata. Panels (c) and (d) give the same views for
\op{WriteEffect}. Rows use the configuration IDs in
Table~\ref{tab:configs}; points give adoption rates and horizontal
whiskers give the reported 95\% intervals. In panels (a) and (c),
the values at right give percentage rate and adopted/evaluable runs/clusters.
In panels (b) and (d), they give rate and
evaluable/stratum-total runs/clusters; the difference between those
run counts is undecidable adoption. Empty strata have no point and
are marked $0/0/0$. Target-value \op{ItemOption} controls are
excluded from the primary analysis. Cluster-bootstrap intervals at an all-zero or all-one boundary describe
only the observed clusters and do not imply population certainty.

For the co-delivered operators, every evaluable run was pre-arrival:
adoption was 100\% for \op{ReturnDate} (149/149) and 59\% for
\op{PaySource} (89/150). For \op{ItemOption}, 47/58 evaluable non-control
runs adopted the fault, including 15/25 pre-arrival and 32/33 late-arrival
runs. Under the stricter literal admission rule, pre-arrival adoption changed
to 6/13, while late-arrival adoption was 16/16; the other sensitivity
analyses did not remove the separation between the two populations.
The difference between aggregate adoption and claim-specific support was
most visible for \op{WriteEffect}. V and T both adopted in every evaluable
run, yet V had no pre-arrival execution and T had only 2/25. Among the same
141 evaluable r0 executions, 101 adopted the fault. Of these, 44 eventually
received counterevidence, but only 17 received it before first use; 27
received it afterward and 57 received none in the observed execution.
The paired checks changed adoption in the expected direction. Removing the
read cue increased \op{WriteEffect} adoption from 76/115 to 115/115.
Making the authoritative source consistent with the injected assertion
changed adoption from 22/25 to 25/25 \op{ItemOption} pairs and from
47/56 to 56/56 \op{WriteEffect} pairs.

\finding{\textbf{A well-defined aggregate adoption result could coexist
with little or no empirical support for a timely-counterevidence claim.}
Aggregation pooled executions with different evidential preconditions, so
the denominator supporting fault propagation was not the denominator
supporting adoption after pre-use counterevidence.}

\subsection{RQ3: Arrival Relative to Faulty Reliance}
\label{sec:rq3}

\begin{figure}[t]
\centering
\includegraphics[width=0.95\linewidth]{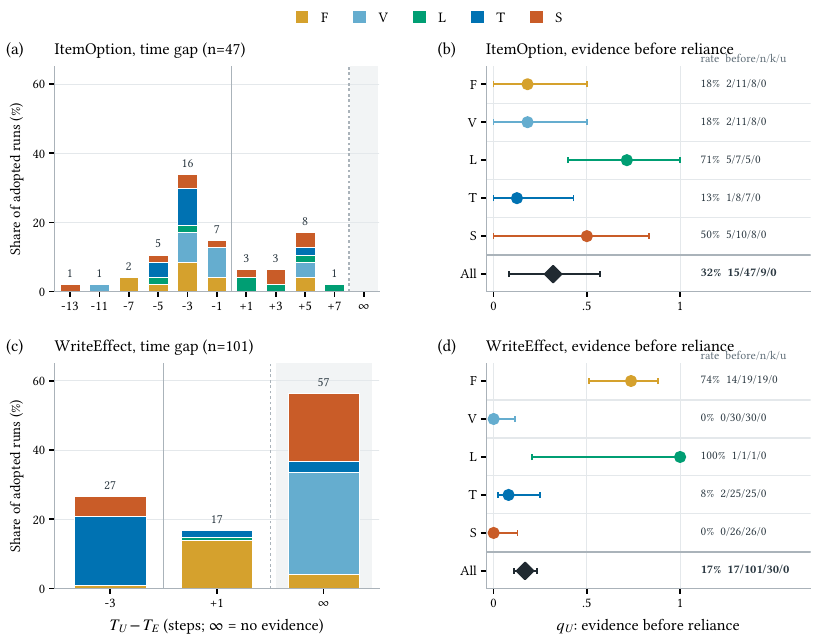}
\caption{Counterevidence arrival relative to faulty reliance.}
\label{fig:rq3-reliance}
\end{figure}

Figure~\ref{fig:rq3-reliance} separates the timing of counterevidence from
the rate at which it preceded first faulty reliance. The rows show
\op{ItemOption} and \op{WriteEffect}. Panels (a) and (c) show the signed gap
$T_U-T_E$ among adopted runs: positive values mean that evidence arrived
before reliance, negative values mean it arrived afterward, and $\infty$
marks evidence absent from the observed execution. Bar labels give run counts;
bar heights give shares of all adopted runs for that operator. Panels (b) and
(d) show $q_U$ by agent configuration, with 95\% intervals and a pooled
diamond. Colors identify the same configurations in the stacked bars and
interval estimates. The right-hand labels give the rate and
before/$n_q$/$k_q$/$u_q$, where $k_q$ counts entities and $u_q$ counts adopted
runs with undecidable ordering.

For the co-delivered operators, counterevidence preceded first faulty
reliance in all 149 adopted \op{ReturnDate} runs and all 89 adopted
\op{PaySource} runs with decidable timing. In Figure~\ref{fig:rq3-reliance}(a),
\op{ItemOption} has 15 positive and 32 negative gaps; panel (b) shows that
the corresponding pooled $q_U$ is 15/47, with configuration estimates from
13\% to 71\%. $T_S=T_U$ in 45/47; the remaining two had
$T_S<T_U$, but counterevidence was already visible before $T_S$.
Adding target-value controls changed $q_U$ by at most three percentage points.

For \op{WriteEffect}, $T_S=T_U$ in all 101 evaluable adopted r0 runs.
Figure~\ref{fig:rq3-reliance}(c) shows 17 positive gaps, 27 negative gaps,
and 57 absent observations; panel (d) gives the pooled $q_U=17/101$.
Configuration estimates span 0\% to 100\%, but the 100\% estimate rests on
one adopted L run. Across both operators, no
primary non-control run had counterevidence arrive between the two use
anchors ($T_S<T_E<T_U$). Thus, the studied operators provide little empirical separation between
the pre-use and pre-reliance conditions; we do not infer from these data
that the two anchors are interchangeable in other execution structures.

First-error timing and final stance did not collapse to the same outcome.
Among adopted \op{ItemOption} late-arrival runs, 20/32 still endorsed the
fault at the end of observation. For \op{WriteEffect}, 14/17 runs with
counterevidence before the first error still ended by endorsing the fault,
whereas 26/27 late-arrival runs no longer endorsed it. Terminal stance does
not establish repair of an earlier external effect. 

\finding{\textbf{The first-error state and the terminal stance were not
interchangeable measures of recovery.}
First dependent use and first faulty reliance usually coincided in the
studied operators, yet later endorsement could diverge sharply from the
first-error state, separating error prevention from subsequent stance
revision and from repair of prior effects.}

\subsection{RQ4: Measurement Recoverability}
\label{sec:rq4}

Table~\ref{tab:rq4} evaluates seven oracle settings against the
human-adjudicated reference. Adoption metrics use applicable faulted runs;
clean runs contribute only to false-positive evaluation. Step accuracy uses
all adopted runs with decidable human $T_U$, counting abstentions as
nonmatches. Panel (a) reports
adoption identification, panel (b) first-faulty-reliance localization, and
panel (c) adoption F1 / exact-step accuracy by operator.

\begin{table*}[t]
\centering
\footnotesize
\setlength{\tabcolsep}{5pt}
\caption{Automated recovery of measurement targets.}
\label{tab:rq4}

\begin{tabular*}{\textwidth}{@{\extracolsep{\fill}}l r r r r r r r@{}}
\toprule
\multicolumn{8}{@{}l}{\textbf{(a) Adoption identification}}\\
Setting & $N_F$ & Coverage & Precision & Recall & F1 &
FP$_{\rm faulted}$ & FP$_{\rm clean}$\\
\midrule
B1 & 120 & 100\% & -- & 0\% & .00 & 0/32 & 0/40\\
B2-L0 & 120 & 87\% & 75\% & 56\% & .64 & 15/22 & 3/39\\
B2-L1 & 120 & 87\% & 81\% & 83\% & .82 & 16/20 & 5/32\\
B3-native & 40 & 98\% & 76\% & 100\% & .87 & 9/10 & 9/10\\
B3-SAA & 120 & 98\% & 82\% & 82\% & .82 & 16/30 & 8/39\\
B5 & 120 & 100\% & 72\% & 84\% & .77 & 29/32 & 0/40\\
\bottomrule
\end{tabular*}

\smallskip
\begin{tabular*}{\textwidth}{@{\extracolsep{\fill}}l r r r r@{}}
\multicolumn{5}{@{}l}{\textbf{(b) First faulty reliance}}\\
Setting & $N_{\rm step}$ & Coverage & Exact step & $\pm1$ step\\
\midrule
B1 & 88 & 0\% & 0\% & 0\%\\
B2-L0 & 88 & 52\% & 34\% & 34\%\\
B2-L1 & 88 & 80\% & 49\% & 49\%\\
B3-native & 30 & 97\% & 43\% & 43\%\\
B3-SAA & 88 & 81\% & 51\% & 51\%\\
B4 & 88 & 100\% & 47\% & 47\%\\
B5 & 88 & 84\% & 0\% & 51\%\\
\bottomrule
\end{tabular*}

\smallskip
\begin{tabular*}{\textwidth}{@{\extracolsep{\fill}}l c c c c@{}}
\multicolumn{5}{@{}l}{\textbf{(c) By operator, adoption F1 / exact step}}\\
Setting & RD & IO & PS & WE\\
\midrule
B1 & .00/0\% & .00/0\% & .00/0\% & .00/0\%\\
B2-L0 & .33/0\% & .79/40\% & .62/60\% & .67/33\%\\
B2-L1 & .95/30\% & .82/43\% & .80/90\% & .69/33\%\\
B3-native & --/-- & .87/43\% & --/-- & --/--\\
B3-SAA & .88/10\% & .81/47\% & .82/90\% & .76/61\%\\
B4 & --/20\% & --/23\% & --/80\% & --/78\%\\
B5 & 1.00/0\% & .79/0\% & .80/0\% & .50/0\%\\
\bottomrule
\end{tabular*}
\end{table*}

Adding the fault-specific relation to B2 increased adoption F1 from .64 to
.82 and exact $T_U$ localization from 34\% to 49\%. B3-SAA likewise reached
.82 adoption F1 but localized $T_U$ exactly in only 45/88 applicable adopted
runs (51\%). Thus, improved adoption identification did not remove the
verdict--localization gap.
Faulted non-adopters exposed a distinct construct error. B2-L1 marked 16/20,
B3-SAA 16/30, and B5 29/32 covered faulted non-adopters as positive, compared
with 5/32, 8/39, and 0/40 false positives on clean controls. Detecting a
fault or anomalous state therefore did not by itself recover faulty reliance.
Errors were also strongly stratum-dependent. B3-SAA adoption recall was
93\%, 90\%, and 9\% in human pre-, late-, and absent-arrival runs,
respectively; exact $T_U$ localization was 61\%, 48\%, and 9\%.
B5 illustrates the temporal distinction: it returned a step in 74/88
applicable adopted runs but matched $T_U$ exactly in none, despite 51\%
$\pm1$ accuracy. Its invariant alarm was often close to, but not identical
to, the semantic first-reliance event.

\finding{\textbf{Aggregate oracle accuracy concealed construct-specific
measurement errors.}
Generic trajectory diagnostics could identify faults or adoption-like
signals without recovering the semantic first-reliance event, and these
errors were uneven across evidential conditions.}

%% file: sections/07_discussion.tex
\section{Discussion}
\label{sec:discussion}

\textbf{Measure Evidential Conditions per Execution.}
Case admission establishes that a timely refutation is possible under the
fixed fault context; it does not assign an evidential condition to subsequent
executions. That condition depends on the observation path realized before
the affected fact is used. Consequently, revealability should remain a
case-level qualification, while arrival should be reconstructed from each
evaluated trajectory.
This separation also clarifies absent arrival. For an admitted case, absence
means that valid counterevidence did not become model-visible in the observed
execution; it does not mean that the interface lacked a legal pre-use
refutation route. Assigning the case-level witness to every run would
therefore erase execution-dependent variation that is part of the evaluation.

\guideline{Use pre-use revealability for case admission, but reconstruct counterevidence arrival from every evaluated execution.}

\textbf{Bind Claims to Their Evidential Denominators.}
Fault propagation, adoption with eventual counterevidence, and adoption under
timely counterevidence require different observed conditions. An aggregate
adoption rate therefore supports a propagation claim over its own evaluable
population; it does not inherit support for a stronger claim whose required
evidential condition is absent from those executions.
Evaluation reports should make this claim--population relation explicit.
When a claim refers to timely counterevidence, its evaluable pre-arrival
population is part of the result. An empty stratum leaves that claim
unobserved, while a small stratum limits the precision with which it can be
characterized, regardless of how well determined the aggregate adoption rate
is.

\guideline{Report each fault-adoption claim together with the evidential
condition it requires and the corresponding evaluable denominator.}

\textbf{Separate Prevention, Revision, and Repair.}
A single recovery label can conflate distinct stages of an execution.
Whether counterevidence precedes the first faulty reliance concerns
\emph{prevention}; whether the agent later stops endorsing the faulty
assertion concerns \emph{revision}; whether an earlier external action or
state change has been undone concerns \emph{repair}. These properties require
different observations.
The two temporal anchors serve complementary roles in this separation.
$T_S$ provides an adoption-independent anchor for the evidential condition,
whereas $T_U$ localizes the first faulty reliance in adopted runs.
$Y_{\mathrm{end}}$ describes a later stance but provides no evidence that
earlier external effects were repaired. Evaluations that study recovery
should therefore state which of these stages their outcome represents.

\guideline{Report error prevention and later stance revision separately; treat repair as a distinct claim that requires evidence about external state.}

\textbf{Validate Automated Measurement Against the Target Construct.}
A trajectory-analysis method is not automatically a valid measurement oracle
because it detects faults, anomalies, or critical steps. The oracle used by
an evaluation must recover the semantic event named by the measurement
construct and, when temporal interpretation is required, its location in the
trajectory.
Validation should therefore cover more than aggregate verdict accuracy.
It should evaluate semantic classification, temporal localization, coverage
or abstention, and error variation across the evidential conditions used in
downstream analysis. Otherwise, oracle errors can change not only individual
labels but also the measured composition of the populations used to support
the evaluation claims.

\guideline{Validate automated oracles separately on the semantic verdict and event location used by the analysis; report coverage and condition-specific errors when those conditions enter downstream claims.}

%% file: sections/08_threats.tex
\section{Threats to Validity}
\label{sec:threats}

\textbf{Measurement validity.}
Case admission and event reconstruction involve semantic judgments about
task relevance, valid counterevidence, dependent use, and faulty reliance.
We mitigate this threat with frozen decision rules, independent review and
adjudication, and validation of the automatic measurements against
human-adjudicated trajectories. The stricter \op{ItemOption} admission
analysis further shows how results change under a narrower interpretation
of task relevance. Our conclusions therefore apply to the constructs and
admission rules stated in this paper.
Arrival and adoption are measured from the same realized trajectories.
Arrival-conditioned comparisons are consequently descriptive rather than
causal. Likewise, terminal stance measures later endorsement of the faulty
assertion, not repair of an earlier external effect. The paired perturbations
support the reported behavioral contrasts but do not isolate counterevidence
arrival as a causal mediator.

\textbf{External validity.}
Our experiments cover two \taubench\ domains, four fault operators, and five
agent configurations. The operators instantiate three counterevidence-
acquisition structures, but each operator realizes a particular task and
interface semantics; other tool contracts or interaction patterns may yield
different arrival distributions and use dynamics. The empirical rates should
therefore be interpreted for the studied settings rather than as population
estimates for tool-using agents in general.
The measurement protocol itself applies to fault-injection settings in which
the affected fact, a task-authoritative refutation, and their relation to
downstream use can be specified from the execution interface. Settings
without such a refutation relation fall outside the scope of the claims
studied here.

%% file: sections/09_related.tex
\section{Related Work}
\label{sec:related}

Fault injection has increasingly been used to evaluate tool-using agents
under unreliable software environments. AgentChaos injects faults at the
LLM--API response boundary and studies their propagation through agent
executions~\cite{agentchaos}. ToolBench-X evaluates recovery from hazards
spanning tool specifications, invocations, execution, outputs, and
cross-source conflicts~\cite{toolbenchx}. ToolMaze and AgentNoiseBench
likewise exercise agents under unreliable tool interactions
\cite{toolmaze,agentnoisebench}. AgentProp-Bench traces parameter errors and
their downstream propagation through later actions and
responses~\cite{agentprop}. Together, these studies characterize how
controlled perturbations affect task outcomes, agent actions, recovery, and
the propagation of corrupted information.

A related body of work examines how agents acquire and use unreliable or
conflicting information. Yang et al.~\cite{agentstrusttools} study the delivery and adoption
of corrupted tool outputs, while work on inconsistent
tool responses and interactive knowledge conflicts examines agent behavior
when competing information becomes available
\cite{xu2026inconsistent,kcbench}. Roh et al.~\cite{beforereasoning} distinguish
failures that occur before evidence inspection from those that remain after a
relevant read, and ToolFailBench separates skipped tool
use from failures to use returned information~\cite{toolfailbench}.
ContainmentBench further analyzes propagation and recovery at different
execution stages after prompt injection~\cite{containmentbench}. These studies
show that fault propagation, evidence acquisition, evidence use, and
downstream failure need not coincide within an execution.

Fault-based testability studies whether a fault can propagate to an
observation that distinguishes faulty from correct behavior
\cite{voas1992pie,ammann2016introduction,li2017ripr,gorz2023mutation,just2014mutants}. Classical work such as
PIE characterizes the conditions required for a fault to become observable,
and subsequent testing research further studies observability and test-oracle
effectiveness~\cite{barr2015oracle}. These studies establish the role of distinguishing observations
in fault revelation. For tool-using agents, we study a related evaluation
problem in which the distinguishing observation is counterevidence to an
injected assertion and its arrival depends on the agent's subsequent tool
interactions. Our measurement separates whether the fault propagates from
whether adoption occurs after such counterevidence becomes visible before use.

Trajectory-diagnosis methods provide another relevant line of work. AgentRx
and Who\&When identify failure causes or critical steps from agent execution
traces~\cite{agentrx,zhang2025whichagent}, while invariant-based techniques
can expose anomalous execution states or tool results. Such techniques provide
candidate mechanisms for reconstructing execution-level events. We therefore
evaluate whether representative diagnosis outputs can recover fault adoption
and the first faulty-reliance event required by the proposed measurement.

%% file: sections/10_conclusion.tex
\section{Conclusion}
\label{sec:conclusion}

In this paper, we studied the evidential conditions under which
fault-injection outcomes for tool-using agents can be interpreted.
A fault case may provide a valid pre-use refutation opportunity without
fixing the evidential condition realized by a particular execution, and an
aggregate adoption result may therefore support a different claim from
adoption observed under timely counterevidence.
We introduced \bench\ to make these conditions explicit through
fault contracts, pre-use revealability admission, and execution-level
reconstruction. Our experiments show that realized counterevidence exposure
can vary after the fault case and agent configuration are fixed, that
claim-specific evidential populations can be sparse even when aggregate
adoption is well defined, that first-error timing and later recovery describe
different execution stages, and that generic trajectory diagnostics do not
necessarily recover the semantic and temporal events required by the
measurement.
These findings motivate four practices for fault-injection evaluation:
measure evidential conditions per execution, bind claims to their evaluable
evidential populations, distinguish prevention, revision, and repair, and
validate automated oracles against the constructs used by the evaluation.
Our study covers four fault operators, five agent configurations, and
1,191 faulted executions. We hope these results support more precise and
reproducible interpretation of fault-injection evaluations for tool-using
agents.